\documentclass[a4paper,fleqn,usenatbib]{mnras}

\usepackage{newtxtext,newtxmath}

\usepackage[T1]{fontenc}
\usepackage{ae,aecompl}

\usepackage{graphicx}	
\usepackage{amsmath}	
\usepackage{amssymb}	

\title[Collimated outflows from BNS merger remnants]{Collimated outflows from long-lived binary neutron star merger remnants}

\author[Riccardo Ciolfi]{
Riccardo Ciolfi,$^{1,2}$
\\
$^{1}$INAF, Osservatorio Astronomico di Padova, Vicolo dell'Osservatorio 5, I-35122 Padova, Italy\\
$^{2}$INFN, Sezione di Padova, Via Francesco Marzolo 8, I-35131 Padova, Italy\\
}

\date{Accepted XXX. Received YYY; in original form ZZZ}

\pubyear{2020}

\begin{document}
\label{firstpage}
\pagerange{\pageref{firstpage}--\pageref{lastpage}}
\maketitle

\begin{abstract}
\noindent The connection between short gamma-ray bursts (SGRBs) and binary neutron star (BNS) mergers was recently confirmed by the association of GRB\,170817A with the merger event GW170817. 
However, no conclusive indications were obtained on whether the merger remnant that powered the SGRB jet was an accreting black hole (BH) or a long-lived massive neutron star (NS).
Here, we explore the latter case via BNS merger simulations covering up to 250\,ms after merger. 
We report, for the first time in a full merger simulation, the formation of a magnetically-driven collimated outflow along the spin axis of the NS remnant.
For the system at hand, the properties of such an outflow are found largely incompatible with a SGRB jet. With due consideration of the limitations and caveats of our present investigation, our results favour a BH origin for GRB\,170817A and SGRBs in general.  
Even though this conclusion needs to be confirmed by exploring a larger variety of physical conditions, we briefly discuss possible consequences of all SGRB jets being powered by accreting BHs.
\end{abstract}


\section{Introduction}

\noindent The coincident detection of gravitational waves (GWs) from the binary neutron star (BNS) merger event GW170817 and the short gamma-ray burst (SGRB) named GRB\,170817A
provided what is widely considered the long awaited compelling evidence that BNS mergers can power SGRBs \citep{LVC-BNS,LVC-MMA,LVC-GRB,Goldstein2017,Savchenko2017,Troja2017,Hallinan2017,Mooley2018a,Lazzati2018,Lyman2018,Mooley2018b,Ghirlanda2019}. 
The combined analysis of the prompt gamma-ray signal and the subsequent X-ray, optical/IR, and radio afterglows ultimately confirmed that GRB\,170817A was powered by a relativistic jet, in accordance with the accepted paradigm (e.g., \citealt{Piran2004,Kumar2015}). 
Moreover, the jet properties were found consistent with the known class of SGRBs, even though this event was the first to be observed off-axis by $15^{\circ}\!-\!30^{\circ}$ (\citealt{Mooley2018b,Ghirlanda2019} and refs.~therein).

This breakthrough discovery left behind a crucial open question on the nature of the central engine producing the jet.
The most likely outcome of the merger was a metastable hypermassive or supramassive neutron star (NS), which eventually collapsed to a black hole (BH). However, it was not possible to firmly establish whether the collapse occurred before or after the SGRB jet was launched. Therefore, an accreting BH (``BH-disk'' scenario, \citealt{Eichler1989,Narayan1992,Mochkovitch1993}) or a rapidly spinning and strongly magnetized NS (``magnetar'' scenario, \citealt{Zhang2001,Gao2006,Metzger2008}) both represent viable central engines for GRB\,1701817A (see, e.g., \citealt{Ciolfi2018} for a recent review). 
\begin{figure*}
  \centering
  \includegraphics[width=0.95\linewidth]{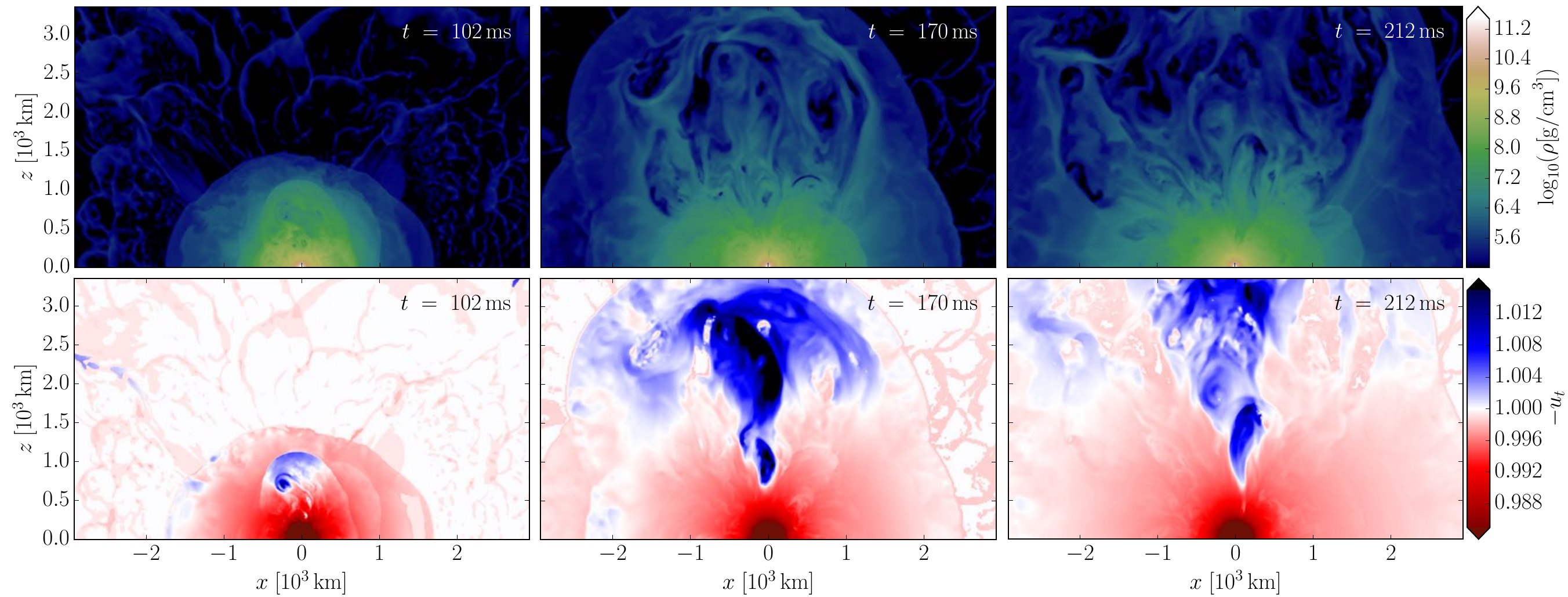}
  \caption{Meridional view of rest-mass density (top) and $-u_t$ (bottom) at 102, 170, and 212\,ms after merger for model `B5e15'. Blue (red) color in the bottom row corresponds to fluid elements that are unbound (bound) according to the geodesic criterion (see, e.g., \citealt{Hotokezaka2013} for $u_t$ definition).}
  \label{fig:evol}
\end{figure*}

Numerical relativity simulations represent the prime tool to investigate how BNS mergers could power SGRB jets. Recent results showed that magnetic fields are most likely the dominant driver of jet formation (e.g., \citealt{Just2016,Ruiz2016}) and thus a proper study of the associated physical mechanisms requires general relativistic magnetohydrodynamic (GRMHD) simulations. 
So far, most simulations of this kind focussed on the BH-disk scenario (e.g., \citealt{Kiuchi2014,Ruiz2016,Kawamura2016}), for which encouraging indications were obtained. In particular, \citet{Ruiz2016} reported the emergence of a mildly relativistic outflow along the BH spin axis, with properties potentially compatible with a nascent SGRB jet.
The alternative scenario, based on a long-lived NS remnant as SGRB central engine, remained instead largely unexplored until \citet{Ciolfi2017} started a first systematic investigation. 
Along the same line, a new simulation extending up to $\sim\!100$\,ms after merger was recently presented in \citet{Ciolfi2019}. No signs of jet formation were found, nor favourable indications that a jet would be launched at later times. 
Although not conclusive, this result reinforced the idea that jet formation might be very challenging without a BH.
\begin{figure}
  \centering
  \includegraphics[width=0.93\linewidth]{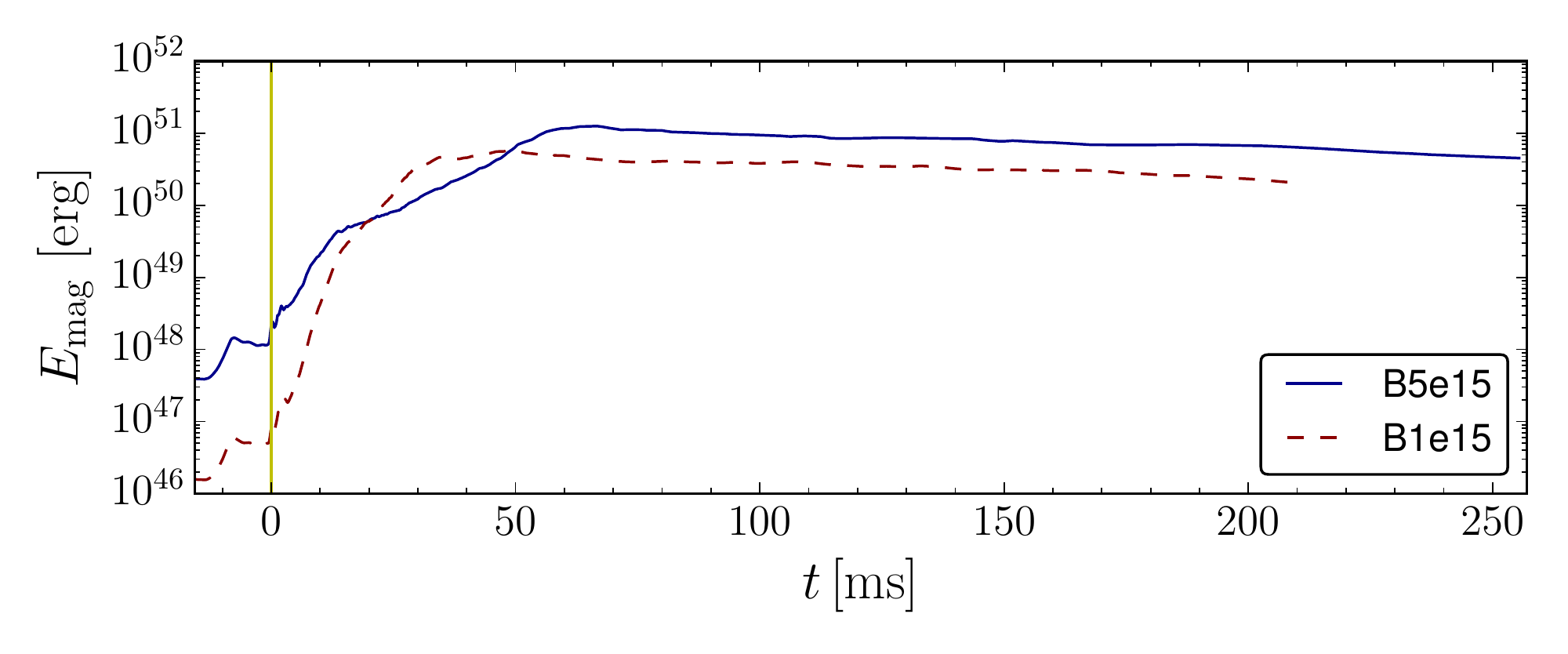}
  \caption{Evolution of total magnetic energy for the two models discussed in this work. Vertical line marks the time of merger.} 
  \label{fig:Emag}
\end{figure}

In this Letter, we present new GRMHD simulations of BNS mergers aimed at exploring the prospects of jet formation from long-lived NS remnants. 
Covering up to more than 250\,ms after merger, these simulations are to date the longest of their kind.
We show, for the first time in a full BNS merger simulation, the formation of a magnetically-driven and collimated outflow in such a system, identifying the launching mechanism and evaluating the associated energetics. 
The BNS model at hand, consistent with the inferred properties of the GW170817 system, offers 
also an opportunity to directly test the long-lived NS central engine hypothesis by assessing whether such an outflow could lead to a SGRB jet compatible with GRB\,170817A.
Our findings reveal a combination of outflow energy, collimation, and Lorentz factor for which producing a SGRB jet appears virtually impossible, 
thus pointing in favour of a BH origin for GRB\,170817A and SGRBs in general. 
We discuss possible consequences that would apply if the above indication is confirmed, as well as caveats and limitations of our current investigation.

\section{Physical model and numerical setup}

\noindent Our initial data reproduce a BNS system with the same chirp mass as estimated for GW170817 \citep{LVC-170817properties} and a mass ratio of $q\!\simeq\!0.9$ (individual masses $\simeq\!1.44,1.29\,M_\odot$). As equation of state (EOS) for NS matter, we adopt a piecewise-polytropic approximation of the APR4 EOS \citep{Akmal:1998:1804} as implemented in \citet{Endrizzi2016}. This choice leads to a merger remnant that does not collapse to a BH within the simulation timespan. As in previous studies (e.g., \citealt{Ciolfi2019} and refs.~therein), the two NSs are endowed with initial dipolar magnetic fields confined to their interior. We perform two different simulations, which only differ for the initial magnetic energy, 
namely $E_\mathrm{mag}\!\simeq\!4\times10^{47}$\,erg and $E_\mathrm{mag}\!\simeq1.6\times10^{46}$\,erg, corresponding to initial maximum field strengths of $5\times10^{15}$~G and $10^{15}$~G, respectively (models `B5e15' and `B1e15'). 
As we discuss in more detail in the next Section, these very high field strengths are chosen to reproduce the magnetization levels expected after merger despite the fact that magnetic field amplification mechanisms are not fully resolved.
\begin{figure*}
  \centering
  \includegraphics[width=0.93\linewidth]{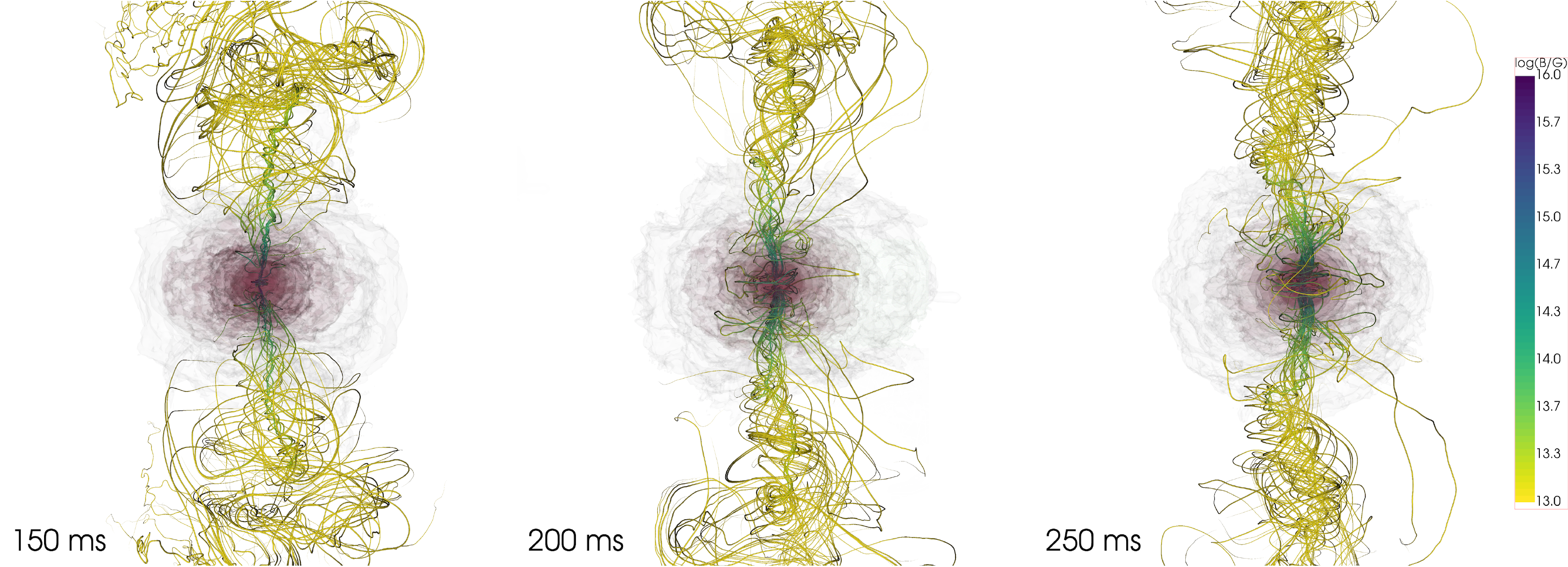}
  \caption{Magnetic field structure 150, 200, and 250\,ms after merger for model `B5e15', with color-coded magnetic field strength. The vertical scale covers $\pm2700$\,km along the $z$-axis. Several semi-transperent isodensity surfaces are also shown for $\rho$ between $6\times10^6$ and $4\times10^9$\,g/cm$^3$ (from light grey to red). } 
  \label{fig:Blines}
\end{figure*}

Numerical codes, methods, and setup are the same adopted in \citet{Ciolfi2017,Ciolfi2019}, with the following exceptions. First, we extended the computational domain up to $\approx\!3400$\,km along all axes. Then, we lowered by two orders of magnitude the level of the artificial floor density ($\rho^*\!\simeq\!6.3\times10^{4}$\,g/cm$^3$), which now corresponds to a mass of $\simeq\!3.5\times10^{-3}\,M_{\odot}$ within a sphere of radius 3000\,km. This ensures that the artificial floor has only a minor impact on the ejecta dynamics. Finally, we employ a resolution with finest grid spacing of $\Delta x\!\approx\!250$\,m.

\section{Evolution and emergence of collimated outflow}

\noindent Among the two models considered, only the one with higher initial magnetic energy leads to the formation of a collimated outflow (Fig.~\ref{fig:evol}), while the other one does not show any sign of it.
The overall merger and post-merger dynamics appear however rather similar up to $\sim\!100$\,ms after merger. 

The evolution of magnetic energy is shown in Fig.~\ref{fig:Emag} for both models. 
At merger, magnetic fields are amplified via the Kelvin-Helmoltz (KH) instability (e.g., \citealt{Kiuchi2015,Kiuchi2018}), developing in the shear layer separating the two NS cores. After few ms, the magnetorotational instability (MRI) \citep{Balbus1991,Duez2006a,Siegel2013} takes over as the dominant source of amplification within the differentially rotating supramassive NS remnant.
The magnetic energy growth saturates at $E_\mathrm{mag}\!\sim\!10^{51}$\,erg, after which a slow and steady decline begins.
Higher initial field strength leads to higher maximum $E_\mathrm{mag}$, but the evolution path is qualitatively different, with $E_\mathrm{mag}$ becoming even higher in the `B1e15' case between $\sim\!20$ and $\sim\!50$\,ms after merger. 
This qualitative difference is indicative of a complex dependence that makes the outcome of other initial magnetizations hard to predict.

We stress that our resolution is not sufficient to fully account for the early small-scale magnetic field amplification (in particular via the KH instability, \citealt{Kiuchi2015,Kiuchi2018}) and therefore, in order to reach the very high magnetization levels expected in the post-merger phase ($E_\mathrm{mag}\!\sim\!10^{51}$\,erg), we imposed initial magnetic energies that are much higher than those of typical merging BNS systems. 
We note, however, that starting from $\sim\!40$\,ms after merger the dominant amplification mechanism (i.e.~the MRI) is well resolved, giving us confidence that the following magnetohydrodynamic evolution is not severely affected by the lack of resolution and that our results are, at least qualitatively, reliable (see analogous case discussed in \citealt{Ciolfi2019}). 

An important aspect common to both models is the absence of an accretion disk around the massive NS remnant. As previously reported in \citet{Ciolfi2019} and confirmed here, if the remnant is strongly magnetized (as expected in real systems) the surrounding matter distribution becomes nearly isotropic within few tens of ms after merger.

In the following, we consider the evolution beyond $100$\,ms after merger, focussing entirely on the `B5e15' case. The first signs that a faster outflow is forming along the remnant spin axis appear soon at $\sim\!600$\,km distance (left panel of Fig.~\ref{fig:evol}). At this time, the baryon mass contained between 100 and 600\,km radial distance from the central object is $\simeq\!0.21\,M_\odot$, while above 600\,km we have $\simeq\!0.02\,M_\odot$. The main front of the post-merger magnetized baryon wind is at $\sim\!1500$\,km.
About $20$\,ms later, the faster material along the axis starts breaking out at $\sim\!1700$\,km, paving the way for a more collimated and even faster incoming outflow.
Within another $\sim\!50$\,ms, the final collimated outflow has emerged, gradually stabilizing towards a conical structure sharply separated from the surrounding much slower baryon wind and entirely contained within a half-opening angle of $15^\circ$ (central and right panels of Fig.~\ref{fig:evol}).
The radial velocity inside this outflow is $0.2-0.3\,c$. 
Finally, towards the end of the simulation ($\sim\!250$\,ms after merger) we observe a significant decline of the outflow power, indicating that the corresponding energy reservoir has been consumed. 

In Fig.~\ref{fig:Blines} we show the magnetic field structure between 150 and 250\,ms after merger, together with representative isodensity surfaces (see \citealt{Ciolfi2019,Kawamura2016} for details on magnetic field line visualization).
Along the spin axis, a jet-like helical structure emerges from the dense environment surrounding the massive NS remnant. As seen for accreting BH systems (e.g., \citealt{Ruiz2016}), this is precisely the type of magnetic field geometry that can accelerate a collimated outflow. 
At the same time, the isodensity surfaces show an isotropic matter distribution around the central object, with no accretion disk. 

\begin{figure}
  \centering
  \includegraphics[width=0.93\linewidth]{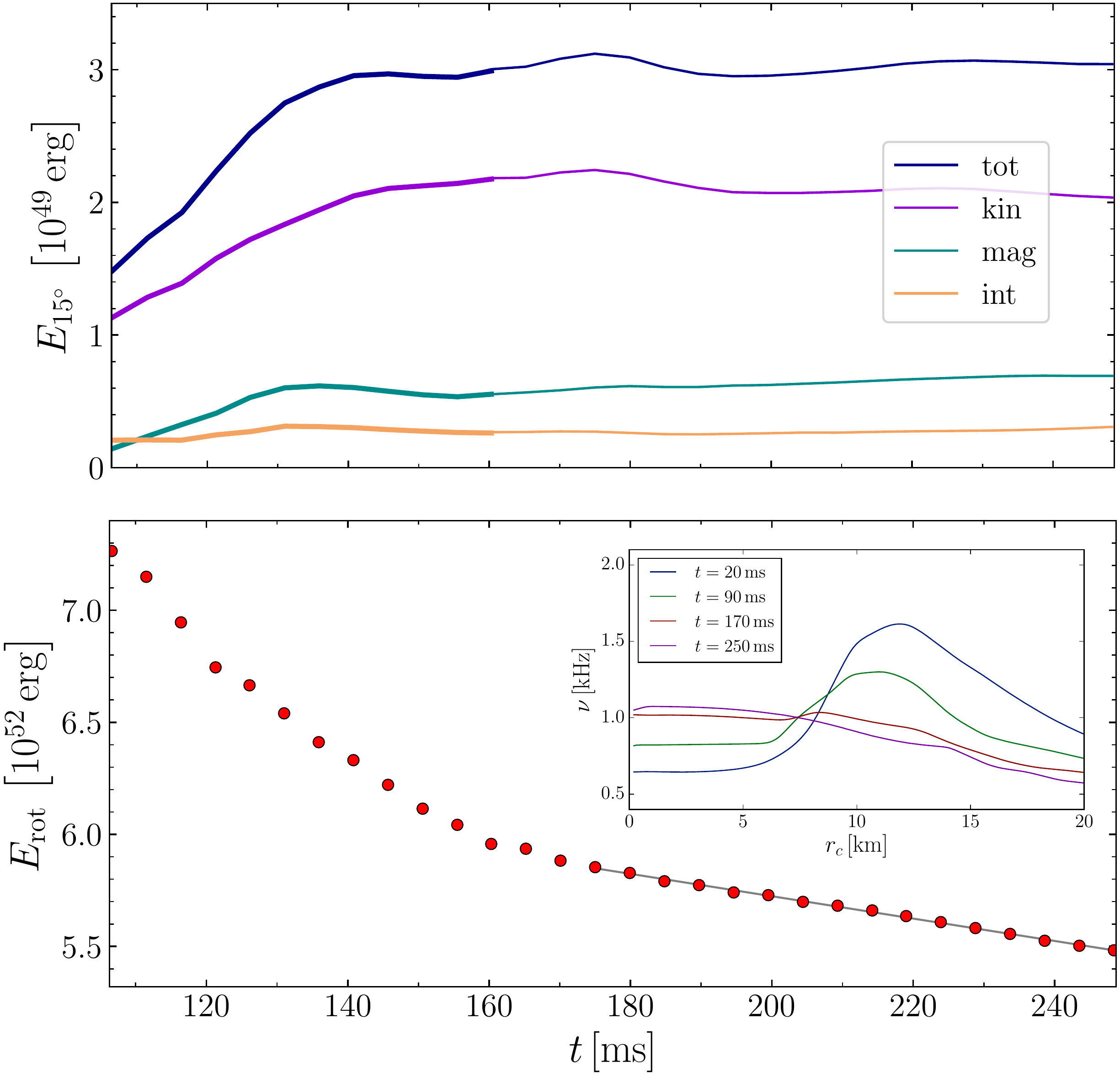}
  \caption{ 
{\it Top}: Evolution of the energy contained within 15$^\circ$ from the $z$-axis and for radial distances larger than 800\,km, given by the sum of radial kinetic energy, magnetic energy, and internal energy. From 160\,ms after merger (thinner line) we include the contribution from the integrated energy flux across the outer boundary of our computational domain. 
{\it Bottom}: Evolution of total rotational energy (red circles), with linear fit for $t\!\geqslant\!175$\,ms after merger (grey line). 
The inset shows the rotation rate versus proper circumferential radius at different times.
} 
  \label{fig:1D}
\end{figure}

We now turn to discuss the energetics of the collimated outflow. In the top panel of Fig.~\ref{fig:1D}, we report the time evolution of the energy contained within $15^\circ$ from the $z$-axis and for radial distances larger than $800$\,km, named $E_{15^\circ}$, given by the sum of radial kinetic energy, magnetic energy, and internal energy. 
Since at $\sim\!160$\,ms after merger the outflow reaches the outer boundary of our computational domain, we also compute the corresponding energy flux and include this contribution in $E_{15^\circ}$. 
The total energy of the collimated outflow, for which $E_{15^\circ}$ represents a very good estimate, grows rapidly up to $\simeq\!3\times10^{49}$\,erg and then flattens with no significant change over the last $\sim\!100$\,ms. This saturation indicates that the energy input from the central engine has significantly declined. 

The total rotational energy of the system is shown in the bottom panel of Fig.~\ref{fig:1D}. The time evolution reveals a drastic change around $170$\,ms after merger, switching from rapid decrease (consistent with an exponential damping) to much slower linear decline ($\dot{E}_\mathrm{rot}\!\simeq\!5\times10^{52}$\,erg/s).
This change perfectly matches the transition from differential to uniform rotation in the core of the remnant NS (cf.~inset of the same panel).
Combined with the flattening of the energy in the collimated outflow, this result provides a strong indication that the energy reservoir powering the latter is given by differential rotation. 
The efficiency in converting rotational energy into outflow energy reaches a maximum of $\eta\!\equiv\!|\dot{E}_{15^\circ}/\dot{E}_\mathrm{rot}|\!\sim\!2.5\times10^{-3}$ at 130\,ms after merger and then it drops rapidly ($<\!10^{-3}$ after 150\,ms). 

Our findings demonstrate that {\it long-lived BNS merger remnants can generate collimated outflows}.
These are launched via a magnetorotational mechanism, where the energy reservoir is given by differential rotation within the remnant NS and part of this energy ($\lesssim\!0.1$\% for the case at hand) is channeled along the spin axis by a strong helical-structured magnetic field.
The acceleration is mainly provided by the radial gradient of magnetic pressure, which is sustained by differential rotation and, in particular along the spin axis, is strong enough to overcome the gravitational pull.
Differently from incipient jets in BH-disk systems, the present outflow is not powered by accretion (no accretion disk is present). 

The above launching mechanism was previously illustrated in GRMHD simulations of idealized, differentially rotating hypermassive or supramassive NSs endowed with an initial poloidal magnetic field imposed by hand (e.g., \citealt{Shibata2011,Kiuchi2012,Siegel2014}). It was shown that the combination of a poloidal magnetic field along the NS spin axis and the differential rotation of the NS itself, without a baryon-polluted surrounding environment, would produce a collimated outflow. Our present simulations reveal that such a mechanism can be successful in actual BNS mergers.

The absence of a collimated outflow in the `B1e15' case shows, however, that not all systems produce a similar outcome. Besides the need for a sufficiently high magnetization, the conditions to form a collimated outflow likely depend on a complex balance between the magnetically driven baryon wind and the emergence of a helical magnetic field structure piercing through it, which are both controlled by the details of magnetic field amplification. This aspect remains poorly understood and requires further investigation. 

\section{Compatibility with the jet of GRB\,170817A}

\noindent The jet core energy of GRB\,170817A as estimated in \citet{Ghirlanda2019} lies in the range $\simeq\!(0.4-29.5)\times10^{49}$\,erg, with central value $\simeq\!4.4\times10^{49}$\,erg. Within the optimistic assumption that all the energy in the emerging collimated outflow ($E_{15^\circ}\!\simeq\!3\times10^{49}$\,erg) will eventually turn into jet core energy, the latter would result compatible with the lower end of the above range. 
Nevertheless, if we consider the isotropic equivalent energy, the estimate for the jet core of GRB\,170817A stands between one and two orders of magnitude above. This is due to the much higher collimation (half-opening angle $\lesssim\!5^\circ$, \citealt{Mooley2018b,Ghirlanda2019}), corresponding to 
$\gtrsim\!10$ times smaller solid angle.
Further collimation of the latter along its propagation might be possible (e.g., via magnetic fields), but reconciling our result with the inferred energy of GRB\,170817A would be in the best case extremely challenging and for most of the allowed range simply impossible.

The discrepancy in Lorentz factor is however what poses the strongest challenge. 
The outflow velocity $3400$\,km away from the remnant is $\lesssim\!0.3\,c$, corresponding to a Lorentz factor $\Gamma\!\lesssim\!1.05$, while a SGRB jet (including the case of GRB\,170817A, \citealt{Mooley2018b,Ghirlanda2019}) would require $\Gamma\!\gtrsim\!10$ or a velocity $\gtrsim\!0.995\,c$. 
In principle, the outflow could still accelerate while propagating at larger distances. 
Nevertheless, the energy-to-mass flux ratio is much smaller than unity ($<\!0.01$) and this excludes any possibility to reach significantly larger Lorentz factors. In other words, the outflow is by far (at least a factor $10^3$) too heavy to be accelerated to $\Gamma\!\sim\!10$ or more.
We conclude that {\it the system under consideration cannot produce a SGRB jet compatible with GRB\,170817A}. 

\section{Consequences for GRB\,170817A and SGRBs in general}

\noindent While we consider here a single combination of EOS and mass ratio, quantitative differences may be expected for other combinations among those compatible with GW170817 and a long-lived NS remnant. 
We note, however, that these differences may not be sufficient to fill the huge gap between the outflow properties we found and those required to explain GRB\,170817A. 
With due caveats (see below), our current results suggest that a scenario in which the jet of GRB\,170817A was powered by a long-lived NS remnant is unlikely, favouring instead a BH-disk central engine as the leading alternative. 

A direct consequence of the above 
conclusion, if confirmed, would be that the massive NS remnant collapsed to a BH in less than $\approx\!1.74$\,s, i.e.~the estimated time separation between merger and onset of gamma-ray emission \citep{LVC-GRB}. This 
would further limit the range of remnant properties consistent with the event and may help in placing additional constraints on the NS EOS (e.g., \citealt{LVC-170817properties,Kastaun2019}). 

Indications disfavouring a long-lived NS remnant as the central engine for GRB\,170817A also cast doubts on the viability of such a scenario for SGRBs in general, or at least for those having comparable or higher jet core energies.
This ``magnetar'' scenario is often invoked as the leading explanation for SGRBs accompanied by long-lasting X-ray plateaus \citep{Rowlinson2010,Rowlinson2013,Lu2015}, 
since the sustained energy injection from long-lived NS remnants can naturally power these features. 
Conversely, if all known SGRBs are powered by BH-disk systems, an alternative scenario to explain the X-ray plateaus 
would become necessary (e.g., the ``time-reversal'' scenario, \citealt{Ciolfi2015,Ciolfi2018}; see also \citealt{Oganesyan2019} and refs.~therein). 

We also note that the overall dynamics found in our simulations would still be relevant within a paradigm where SGRB jets are always launched after BH formation. In this case, the emergence of a collimated outflow like the one discussed here might often precede the actual SGRB jet production, with potential impact on the final jet properties (e.g.,~the angular structure). Moreover, the presence of such an outflow could lead to detectable precursor signals \citep{Troja2010} (e.g., produced when it breaks out of the nearly isotropic baryon-loaded wind).

\section{Caveats}

\noindent A main caveat accompanying our results is given by the fact that the early magnetohydrodynamic evolution (at least up to $\sim\!40$\,ms after merger) cannot be properly resolved with current computational resources (e.g., \citealt{Kiuchi2015,Kiuchi2018}; see discussion on Fig.~\ref{fig:Emag}).
A much higher resolution would allow the system to reach similar magnetization levels starting from much lower (and more realistic) pre-merger field strengths. At the same time, it would have a quantitative impact on both the magnetically driven baryon wind and the development of a jet-like helical structure along the remnant spin axis.
There is no secure way to predict whether this would make the emergence of a collimated outflow easier or more challenging for a given set of initial models.
However, for those systems where such an outflow is produced, 
the latter might still be too baryon loaded to have any chance of evolving into a SGRB jet.

More general initial magnetic field geometries with poloidal components extending outside the two NSs could also have an impact, possibly facilitating the formation of strong helical fields in the post-merger phase (e.g., \citealt{Ruiz2016}). Whether this could significantly alter our general conclusions remains to be investigated. 

Next, we note that we are neglecting neutrino radiation.
Besides modifying the composition (i.e.~the electron fraction) of the post-merger baryon wind, neutrino emission and reabsorption could further enhance mass ejection, resulting in a larger baryon pollution opposing to the propagation of a collimated outflow. 
Based on this, 
we could argue that even including neutrino radiation, the difficulties in launching SGRB jets without a BH would persist. Nevertheless, this needs to be confirmed with actual simulations combining together magnetic and neutrino effects. 

Finally, the introduction of realistic NS spins before merger could affect the remnant rotational profile (e.g., \citealt{Ruiz2019}). While this might have quantitative effects on the results, we expect that the qualitative outcome (and related conclusions) would most likely remain unchanged. 

\section{Concluding remarks}

\noindent Our results show that long-lived BNS merger remnants can produce collimated outflows, 
launched via a magnetorotational mechanism where differential rotation represents the main energy reservoir. Furthermore, our study reveals outflow properties that cannot be reconciled with a SGRB jet, due to the high baryon content and, consequently, rather low maximum Lorentz factor achievable.
These indications favour a BH origin for SGRB jets. However, we also point out possible caveats and additional effects that might alter such a conclusion. 

If collimated outflows like the one discussed here cannot lead to a SGRB, they could still produce potentially detectable transients that
manifest theirselves as SGRB precursors or, in a fraction of BNS merger events, as independent signals. 
We therefore encourage future investigation to establish the bulk properties that such transients would have and possible links to other known transient populations.

\bigskip

\noindent We thank B.~Giacomazzo and W.~Kastaun for useful discussions. Numerical simulations were performed on the cluster MARCONI at CINECA (Bologna, Italy). We acknowledge a CINECA award under the MoU INAF-CINECA initiative (Grant INA17\_C3A23), for the availability of high performance computing resources and support. In addition, part of the numerical calculations have been made possible through a CINECA-INFN agreement, providing access to further resources (allocation INF19\_teongrav).



\bibliographystyle{mnras}
\bibliography{refs}



\bsp	
\label{lastpage}
\end{document}